\documentclass[prb]{revtex4}
\input{psfig.sty}
\usepackage{epsfig}
\usepackage{dcolumn}
\usepackage{amsmath}
\usepackage{amsmath}
\usepackage{bbm}
\usepackage{graphicx}

\begin{document}
\title{\bf Spin filtering through excited states in double quantum dot pumps}
\author{Rafael S\'anchez$^1$, Ernesto Cota$^2$, Ram\'on Aguado$^1$ and Gloria Platero$^1$}
\affiliation{ 1-Instituto de Ciencia de Materiales, CSIC, Cantoblanco, Madrid,28049, Spain. \\
2-Centro de Ciencias de la Materia Condensada - UNAM, Ensenada, Mexico.\\ }
\date{\today}
\begin{abstract}
Recently it has been shown that ac-driven double quantum dots can
act as spin pumps and spin filters. By calculating the current
through the system for each spin polarization, by means of the
time evolution of the reduced density matrix in the sequential
tunneling regime (Born-Markov approximation), we demonstrate that
the spin polarization of the current can be controlled by tuning
the parameters (amplitude and frequency) of the ac field.
Importantly, the pumped current as a function of the applied
frequency presents a series of peaks which are uniquely associated
with a definite spin polarization. We discuss how excited states
participating in the current allow the system to behave as a
bipolar spin filter by tuning the ac frequency and intensity. We
also discuss spin relaxation and decoherence effects in the pumped
current and show that measuring the width of the current vs
frequency peaks allows to determine the spin decoherence time
$T_{2}$.
\end{abstract}
\maketitle
\section{Introduction}
Production, control and detection of spin polarized current
through nanostructures ({\it spintronics})\cite{spintronics} has become
an area of intense activity in the past few years. This is mostly
due to the long coherence times of spins, as compared to charge,
giving rise to possible applications in quantum information
processing. Proposals for generating spin-polarized currents
include spin injection by using ferromagnetic metals \cite{ferro}
or magnetic semiconductors \cite{semimag}. Alternatively, one may
use quantum dots (QDs) as spin filters or spin pumps
\cite{spin-pump1,aono,cotanano,cotaPRL}. In a semiconductor QD, the number of
electrons can be controlled via gate voltages, making it an ideal
system to study transport of individual electrons and probe the
structure and properties of the discrete energy level spectrum.
Indeed, the spin of an isolated electron in a QD has been proposed
as a quantum bit for transport of quantum information. For QD spin
filters, dc transport through few electron states is used to
obtain spin-polarized currents which are almost 100\%
spin-polarized as demonstrated experimentally by Hanson {\it et
al} \cite{hanson} following the proposal of Recher {\it et al}
\cite{recher}. The basic principle of spin pumps is closely
related to that of charge pumps. In a charge pump a dc current is
generated by combining ac driving with either absence of inversion
symmetry in the device, or lack of time-reversal symmetry in the
ac signal. The range of possible pumps includes turnstiles \cite{miriam},
adiabatic pumps or non-adiabatic pumps based on photon-assisted
tunneling (PAT) \cite{report1,report2}. In the last few years, the
application of ac electric fields in quantum dots has shown to be
very accurate both to control and to modify their transport and
electronic properties\cite{report1}. For instance, Sun et al.\cite{sun}
have proposed a spin cell based on a
double quantum dot (DQD) driven by microwave radiation in the
presence of an external {\it non-uniform} magnetic field.
For strongly
correlated electrons in quantum dots in the Kondo regime, the
application of an ac potential has been shown to induce decoherence and destroy
the Kondo peak\cite{rosikondo, elzerman}. An ac potential modifies
as well the electron dynamics within single and double quantum
dots and it has been shown that under certain conditions it
induces charge localization and destroys the tunneling, allowing,
by tuning the parameters of the ac field, to control the time
evolution of the state occupation, and therefore the entanglement
character of the electronic wave function\cite{charles, report1}.
Regarding detection of single spin states in semiconductor quantum dots, this has been
achieved\cite{single-shot,hanson3} using quantum point
contacts and spin to charge conversion. These studies allow determination of spin relaxation times on
the order of milliseconds.

Recently, we proposed a new scheme for realizing {\it both} spin
filtering and spin pumping in an ac-driven DQD (PAT regime), in
the Coulomb blockade transport regime, by tuning the parameters of
the ac field\cite{cotaPRL}. We considered an asymmetric DQD system
with one energy state in the left dot and two orbital states in
the right dot, allowing each QD to contain up to two electrons. An
in-plane magnetic field breaks the spin degeneracy producing
Zeeman splitting. Thus, for the right dot, we can have the
following singlet and triplet states, in order of increasing
energy:
$|S_{0}\rangle=\frac{1}{\sqrt2}\left(|\downarrow\uparrow\rangle-|\uparrow\downarrow\rangle\right)$,
$|T_{+}\rangle=|\uparrow\uparrow^*\rangle$,
$|T_{0}\rangle=\frac{1}{\sqrt2}\left(|\downarrow\uparrow^*\rangle+|\uparrow\downarrow^*\rangle\right)$,
$|T_{-}\rangle=|\downarrow\downarrow^*\rangle$ and
$|S_{1}\rangle=\frac{1}{\sqrt2}\left(|\downarrow\uparrow^*\rangle-|\uparrow\downarrow^*\rangle\right)$,
where the electrons in the upper level are marked with an asterisk
($^*$).  In our previous work\cite{cotaPRL}, we considered a basis
up to the lowest energy triplet state $|T_+\rangle$, and we
discussed the conditions for obtaining spin polarized current from
unpolarized leads. Here, we complement these results by adding the
remaining excited triplet states which also participate in the
pumping process. We find that at certain frequency of the ac field
and in the presence of an uniform magnetic field the double
quantum dot acts as a pump of fully spin polarized electrons.
Increasing the ac frequency, we show that an inhomogeneous
magnetic field is necessary in order to obtain fully spin
polarized current in the reverse direction. Then the system, in
such a configuration, acts as a {\it bipolar} spin filtering just
by tuning the ac frequency.

The periodic variation of the gate potentials in our system,
consisting on an asymmetric DQD, allows for a net dc current
through the device even with no dc voltage applied
\cite{stafford,hazelzet}: if the system is driven at frequencies
(or subharmonics) corresponding to the energy difference between
two time-independent eigenstates related by the intradot tunneling
of an electron, this electron become completely delocalized
\cite{oosterkamp,petta}. If the left reservoir (chemical potential
$\mu_L$) can donate electrons to the left dot (at a rate
$\Gamma_L$) and the right reservoir (chemical potential $\mu_R$)
can accept electrons from the right dot (at a rate $\Gamma_R$) the
system will then pump electrons from left to right, even when
there is no dc bias applied, namely $\mu_L=\mu_R$. Starting from
this pumping principle our device has two basic characteristics:
{\it i)} If the process involves two-particle states, the pumped
current can be completely spin-polarized {\it even if the contact
leads are not spin polarized} and {\it ii)} the pumping can occur
either through triplet (Fig. 1a) or singlet (Fig. 1b) states
depending on the applied frequency, such that the {\it degree of
spin polarization} can be tuned by means of the ac field. For
example, if one drives the system (initially prepared in a state
with $n=n_L+n_R=3$ electrons: $|L=\downarrow\uparrow,R=\uparrow
\rangle$) at a frequency corresponding to the energy difference
between the singlets in both dots, namely
$\omega_{S_0\downarrow}\approx E_{S_0,R}-E_{S_0,L}$ (we consider
$\hbar=e=1$), the electron with spin $\downarrow$ becomes
delocalized in the DQD system. If now the chemical potential for
taking $\downarrow$ ($\uparrow$) electrons out of the right dot is
above (below) $\mu_R$, a spin-polarized current is generated. The
above conditions for the chemical potentials can be achieved by
breaking the spin-degeneracy through a Zeeman term in each dot $\Delta_\alpha =
|g|\mu_B B_\alpha$ ($\alpha=L,R$), where $B_\alpha$ is the external magnetic field at dot $\alpha$,  (which is
applied parallel to the sample in order to minimize orbital
effects), $g$ is the effective g-factor and $\mu_B$ the Bohr
magneton. Then, for example, a fully spin-down polarized pump is
realized at the frequency $\omega_{S_0\downarrow}$ through the sequence: $(\downarrow\uparrow,\uparrow
0)\stackrel{\rm AC}\Leftrightarrow (\uparrow,\downarrow\uparrow 0)
\stackrel{\Gamma_R}\Rightarrow (\uparrow,\uparrow
0)\stackrel{\Gamma_L}\Rightarrow (\downarrow\uparrow,\uparrow 0)$
or $(\downarrow\uparrow,\uparrow 0)\stackrel{\rm
AC}\Leftrightarrow (\uparrow,\downarrow\uparrow 0)
\stackrel{\Gamma_L}\Rightarrow
(\downarrow\uparrow,\downarrow\uparrow 0
)\stackrel{\Gamma_R}\Rightarrow (\downarrow\uparrow,\uparrow 0)$
which involves singlet states. Increasing the frequency of the ac
voltage, the states $(\downarrow\uparrow,\uparrow 0)$ and
$(\downarrow,T_+)$ are brought into resonance at
$\omega_{T_+\uparrow}\approx E_{T_+,R}-E_{S_0,L}$ and spin up
polarized current flows to the collector from the right quantum
dot. However, it can be seen that the
state $(\downarrow\uparrow,\uparrow 0)$ is also in resonance with
$(\uparrow,T_0)$ at the frequency
$\omega_{T_0\downarrow}\approx
E_{T_0,R}-E_{S_0,L}=\omega_{T_{+\uparrow}}+\Delta_R-\Delta_L$, which coincides with $\omega_{T_{+\uparrow}}$ for an homogeneous magnetic
field,
$\Delta_R=\Delta_L$, and
this process contributes to spin down current. Consequently, the
current at this frequency is partially spin polarized. This
degeneracy, present in the three electron configuration, is
removed if the Zeeman splitting is different in the left and right
dot. This could experimentally be achieved either by applying a
different magnetic field to each dot or considering quantum dots
with different g-factors. In that case, $\omega_{T_{+\uparrow}}
\ne \omega_{T_0\downarrow}$ and a pure spin up current is obtained
by applying an ac potential with frequency $\omega_{T_+\uparrow}$.

In practical implementations of such devices, the effects of spin
relaxation and decoherence, characterized respectively by times
$T_1$ and $T_2$, need to be addressed. The most important source
of decoherence in quantum dots is the hyperfine interaction
between electron and nuclear spins.  Recently, several studies
have been reported \cite{ono,koppens,pettaScience} measuring these
characteristic times and also ${T_2}^*$, the spin dephasing time
for an ensemble of nuclear spins. Decoherence effects are included
in our model in a phenomenological way and we give an analytical
treatment to explain the numerical results. Interestingly, we find
that the decoherence time $T_2$ can be obtained from the width of
the current vs frequency peaks.


This paper is organized as follows: in section II we introduce our
model and write the equations for the time evolution of the
density matrix. In section III we describe the results obtained
for homogeneous and inhomogeneous magnetic field. In section IV we
study spin relaxation effects both numerically and analytically,
and discuss the possibility of obtaining the decoherence time
$T_2$ from the width and height of the current vs frequency peaks.
We finalize with conclusions and outlook in section V.

\section{Theoretical model}
Our system consists of an asymmetric DQD
connected to two
Fermi-liquid leads which are in equilibrium
with
reservoirs kept at the chemical potentials $\mu_\alpha$,
$\alpha=L,R$.  Using a standard tunneling Hamiltonian approach, we
write for the full Hamiltonian
$\mathcal{H}_l+\mathcal{H}_{DQD}+\mathcal{H}_Z+\mathcal{H}_T$, where
$\mathcal{H}_l=\sum_{\alpha}\sum_{k_\alpha,\sigma}
\epsilon_{k_\alpha} c_{k_\alpha\sigma}^\dagger c_{k_\alpha\sigma}$
describes the leads and
$\mathcal{H}_{DQD}=\mathcal{H}_{QD}^L+\mathcal{H}_{QD}^R+\mathcal{H}_{L\Leftrightarrow
R}$ describes the DQD.
$\mathcal{H}_{QD}^L,\mathcal{H}_{QD}^R$ describe each dot
including the charging energies of the dot electrons. The presence of an external magnetic field is taken into account through the term
$\mathcal{H}_Z=\frac{1}{2}\sum_{\alpha,\sigma,\sigma'} \Delta_{\alpha}
d_{\alpha,\sigma}^\dagger (\sigma_z)_{\sigma\sigma'}d_{\alpha,\sigma'}$, where $\Delta_\alpha\equiv g\mu_B B_\alpha$ is the
Zeeman splitting of the energy levels of each QD in the
presence of an external magnetic field ${\vec B_\alpha}=(0,0,B_\alpha)$.
The model has been described in detail elsewhere \cite{cotaPRL} but we include the description here for the sake of clarity.
It is assumed that only one orbital
in the left dot participates in the spin-polarized pumping process
whereas \emph{two} orbitals in the right dot (energy separation
$\Delta\epsilon$) are considered. The isolated left dot is thus
modelled as a one--level Anderson impurity:
$\mathcal{H}_{QD}^L=\sum_{\sigma} E_L^\sigma d^\dagger_{L\sigma}
d_{L\sigma}+ U_L n_{L\uparrow} n_{L\downarrow}$, whereas the
isolated right dot is modelled as: $\mathcal{H}_{QD}^R = \sum_{i
\sigma} E_{R i}^\sigma d^\dagger_{R i\sigma} d_{Ri \sigma} +
U_R(\sum_i n_{R i\uparrow} n_{R
i\downarrow}+\sum_{\sigma,\sigma'}n_{R 0\sigma} n_{R 1\sigma'})+ J
{\bf S}_0  {\bf S}_1$ including the exchange interaction term. The
index $i=0,1$ denotes the two levels. In practice, we take
$E_L^\uparrow=E_{R 0}^\uparrow=0$, $E_L^\downarrow=\Delta_L$ and
$E_{R 0}^\downarrow=\Delta_R$, with charging energies $U_R>U_L$.
${\bf S}_i = (1/2)\sum_{\sigma \sigma'} d^\dagger_{R i\sigma} {\bf
\sigma}_{\sigma \sigma'} d_{R i\sigma'}$ are the spin operators of
the two levels.
As a consequence of Hund's rule, the intra--dot exchange, $J$, is
ferromagnetic ($J < 0$) such that the energy of the singlet
$|S_1\rangle = (1/\sqrt{2}) (d^\dagger_{R0\uparrow}
d^\dagger_{R1\downarrow} - d^\dagger_{R0\downarrow}
d^\dagger_{R1\uparrow}) |0\rangle$
($E_{S_1,R}=U_R+\Delta_R+\Delta\varepsilon-\frac{3J}{4}$) is
higher than the energy of the triplets $|T_+\rangle =
d^\dagger_{R0\uparrow} d^\dagger_{R1\uparrow} |0\rangle$,
($E_{T_+,R}=U_R+\Delta\varepsilon+\frac{J}{4}$), $|T_0\rangle =
(1/\sqrt{2}) (d^\dagger_{R0\uparrow} d^\dagger_{R1\downarrow} +
d^\dagger_{R0\downarrow} d^\dagger_{R1\uparrow}) |0\rangle$
($E_{T_0,R}=U_R+\Delta_R+\Delta\varepsilon+\frac{J}{4}$), and
$|T_-\rangle = d^\dagger_{R0\downarrow}
d^\dagger_{R1\downarrow}|0\rangle$
($E_{T_{-1},R}=U_R+2\Delta_R+\Delta\varepsilon+\frac{J}{4}$). As
can be seen, due to the Zeeman splitting, $E_{T_-} >
E_{T_0}>E_{T_+}$. Finally, we consider the case where
$\Delta\epsilon >\Delta_R-J/4$ such that the triplet $|T_+\rangle$
is higher in energy than the singlet $|S_0\rangle = (1/\sqrt{2})
(d^\dagger_{R0\uparrow} d^\dagger_{R0\downarrow} -
d^\dagger_{R0\downarrow} d^\dagger_{R0\uparrow}) |0\rangle$
($E_{S_0,L(R)}=U_{L(R)}+\Delta_{L(R)}$).
$\mathcal{H}_{L\Leftrightarrow R}=\sum_{i,\sigma}t_{LR}(
d_{L\sigma}^\dagger d_{R i\sigma} + h.c.)$ describes interdot
tunneling. The tunneling between leads and each QD is described by
the perturbation
$\mathcal{H}_T=\sum_{k_L,\sigma}V_{L}(c_{k_L\sigma}^{\dag}d_{L\sigma}+{\rm
h.c.})+\sum_{i,k_R,\sigma}V_{R}(c_{k_R\sigma}^{\dag}d_{R
i\sigma}+{\rm h.c.})$.
In addition we consider an
external ac field acting on the dots, such that the single
particle energy levels become time dependent:
\begin{equation}
E_{L(R)}^\sigma \rightarrow E_{L(R)}^\sigma(t)=E_{L(R)}^\sigma \pm\frac{V_{AC}}{2}\cos \omega t
\label{AC-eq}
\end{equation}
where $\sigma=\uparrow,\downarrow$, $V_{AC}$ and $\omega$ are the
amplitude and frequency, respectively, of the applied field. We have
considered a basis of 40 states in the particle number
representation which are obtained under these conditions,
considering up to two electrons in each QD. To study the dynamics
of a system connected to reservoirs one can consider the reduced
density matrix (RDM) operator, $\hat\rho=tr_R\hat\chi$, where one
traces all the reservoir states in the complete density operator of
the system, $\hat\chi$. The evolution of the system will be given
by the Liouville equation: $\frac{d\hat\rho(t)}{dt}=-i[\hat
H(t),\hat\rho(t)]$. Assuming the Markov approximation\cite{Blum},
we obtain the master equation, written as:
\begin{eqnarray}
\dot\rho(t)_{m'm}&=&-i\omega_{m'm}(t)\rho_{m'm}(t)-i[\hat H_{L\Leftrightarrow R},\hat\rho(t)]_{m'm} \nonumber\\
&&+\left(\sum_{k\ne m}\Gamma_{mk}\rho_{kk}(t)-\sum_{k\ne m}\Gamma_{km}\rho_{mm}(t)\right)\delta_{m'm} \nonumber\\
&&-\gamma_{m'm}\rho_{m'm}(t)(1-\delta_{m'm})
\label{mastereq}
\end{eqnarray}
where $\omega_{m'm}(t)=E_{m'}(t)-E_m(t)$ is the energy difference between
the states $|m\rangle$ and $|m'\rangle$ of the isolated DQD,
$\Gamma_{mk}$ are the transition rates for electrons tunneling through the leads, from state $|k\rangle$
to state $|m\rangle$ and
$\gamma_{m'm}$ describes the decoherence of the DQD states due to the interaction with the
reservoir. This decoherence rate is related with the transition rates by:
$\Re\gamma_{m'm}=\frac{1}{2}\left(\sum_{k\ne m'}\Gamma_{km'}+\sum_{k\ne m}\Gamma_{km}\right)$.

Neglecting the influence of the ac field on tunneling processes
through the leads, one can calculate these rates by the Fermi
Golden Rule approximation:
\begin{equation}
\Gamma_{mn}=\sum_l \Gamma_l (f(\omega_{mn}-\mu_l)\delta_{N_m,N_n+1}
+(1-f(\omega_{nm}-\mu_l))\delta_{N_m,N_n-1}),
\end{equation}
where $\Gamma_l=2\pi \mathcal D_l
|V_l|^2, l=L,R$ are the tunneling rates for each lead. It is assumed that the
density of states in both leads $\mathcal D_{L,R}$ and the
tunneling couplings $V_{L,R}$ are energy-independent. $N_k$ is the
number of electrons in the system when it is in state $|k\rangle$.

We include spin relaxation and decoherence phenomenologically in
the corresponding elements of the equation for the RDM. Relaxation
processes are described by the spin relaxation time
$T_1=(W_{\uparrow\downarrow}+W_{\downarrow\uparrow})^{-1}$, where
$W_{\uparrow\downarrow}$ and $W_{\downarrow\uparrow}$ are
spin-flip relaxation rates fulfilling a detailed balance:
$W_{\downarrow\uparrow}=
\exp(-\Delta_z/k_BT)W_{\uparrow\downarrow}$, where $k_B$ is the
Boltzman constant and $T$ the temperature. A lower bound for the
spin relaxation time $T_1$ on the order of $\mu{\rm s}$ at $B\approx 0-2T$
was obtained by Fujisawa et al.\cite{fujisawa}. Recently a value
of $T_1=2.58$ ms with a field $B=0.02$ T was measured
\cite{hanson3} for a single QD using a tunnel-rate-selective
readout method.  In the following, we focus on zero temperature
results such that $W_{\downarrow\uparrow}=0$ and thus
$T_1=W_{\uparrow\downarrow}^{-1}$.  $T_2$ is the spin decoherence
time, i.e., the time over which a superposition of opposite spin
states of a single electron remains coherent. Recently, Loss et
al. \cite{Loss-relaxation} obtained that $T_2=2T_1$ for spin
decoherence induced by spin-orbit interaction. This time can be
affected by spin relaxation and by spin dephasing time ${T_2}^*$,
i.e., the spin decoherence time for an ensemble of spins. For
processes involving hyperfine interaction between electron and
nuclear spins, Petta et al.\cite{pettaScience} have obtained
${T_2}^*\approx 10 \,{\rm ns}$
 from singlet-triplet spin relaxation studies in a DQD.
Here we consider two cases: ${T_2}^*$=0.1$T_1$ and ${T_2}^*=0.001T_1$.\\
In practice,
we integrate numerically the dynamics of the RDM in the chosen
basis. In particular, all the results shown in the next paragraphs
are obtained by letting the system evolve from the initial state
$|\downarrow\uparrow,\uparrow\rangle$ until a stationary state is
reached. The dynamical behavior of the system is governed by rates
which depend on the electrochemical potentials of the
corresponding transitions.

We do not include second order processes as co-tunneling.
 The current from the right dot to the right contact is given by:
$I_{L\rightarrow R}(t)=\Gamma_R\sum_{s} \rho_{ss}(t)$,
with a similar expression for $I_{R\rightarrow L}$. Here, states
$|s\rangle$ are such that the right dot is occupied.

\section{Results}
\label{sec:results}
 We analyze the current through an ac-driven
double quantum dot, with $V_{AC}$ and $\omega$ the amplitude and
frequency of the ac-field, weakly coupled to the external leads,
in the presence of a magnetic field which induces a spin splitting
$\Delta_z$ in the discrete states of each dot. We assume the
ground state in each dot to be the one with spin-up. The model
that we propose is of general application, i.e., it is valid for
both small or large inter-dot coupling. In particular we show
results for a configuration where the coupling of the dots with
the leads is weak and symmetric: $\Gamma_{L,R}=0.001$, the hopping
between the dots is $t_{LR}=0.005$ and the charging energy for the left
and right dots are $U_L=1.0$ and $U_R=1.3$ respectively. All
energy units are in $meV$. The Zeeman splitting produced by an
external homogeneous magnetic field is: $\Delta_L=\Delta_R=0.026$,
corresponding to a magnetic field $B\approx 1T$; the exchange
constant for the right dot is $J=-0.2$ and the chemical potentials
in the left and right leads are $\mu_L=\mu_R=1.31$, respectively.
We have considered two levels in the right dot with energy
separation $\Delta\epsilon=0.45$ and we neglect inter-dot Coulomb
and exchange interactions.

In order to configure the system in a way that electrons can
tunnel from the right dot to the right lead only through the
transitions:
$|\uparrow\downarrow\rangle_R\rightarrow|\uparrow\rangle_R$
we choose the energy parameters so that they satisfy:
$U_R<\mu_R<U_R+\Delta_R$.
The energy cost of introducing a second electron with either spin-up or spin-down
polarization in the left dot has to be
smaller than the chemical potential of the left lead. This is always
satisfied if $\mu_L>U_L+\Delta_L$. We consider the system to be in the pumping configuration
$\mu_L=\mu_R$
throughout.

Then, if the DQD is initially in the state
$|\uparrow\downarrow,\uparrow\rangle$, no current will flow
through the system unless the ac frequency fits the energy
difference between the different states for the left and right
dots. For instance:
$|\uparrow\downarrow,\uparrow\rangle\Leftrightarrow|\uparrow,\uparrow\downarrow\rangle$
(at $\omega=\omega_{S_0\downarrow}=U_R-U_L+\Delta_R-\Delta_L)$,
$|\uparrow\downarrow,\uparrow\rangle\Leftrightarrow|\uparrow,T_0\rangle$
(at
$\omega=\omega_{T_0\downarrow}=U_R-U_L+\Delta_R-\Delta_L+\Delta\epsilon+J/4)$,
$|\uparrow\downarrow,\uparrow\rangle\Leftrightarrow|\uparrow,S_1\rangle$
(at
$\omega=\omega_{S_1\downarrow}=U_R-U_L+\Delta_R-\Delta_L+\Delta\epsilon-3J/4)$
or
$|\uparrow\downarrow,\uparrow\rangle\Leftrightarrow|\downarrow,T_+\rangle$
(at $\omega=\omega_{T_+\uparrow}=U_R-U_L+\Delta\epsilon+J/4)$. The
suffix $\uparrow(\downarrow)$ is used to remark that the inter-dot
tunneling electron has spin-up (down) polarization. At these
frequencies, one electron becomes delocalized undergoing Rabi
oscillations with frequency\cite{report1}
\begin{equation}
\Omega_{R}=2t_{LR}J_\nu(\frac{V_{ac}}{\omega}),
\label{Rabi}
\end{equation}
where $J_\nu$ is the $\nu th$-order Bessel function of the first
kind and $\nu$ is the number of photons required for bringing the
two states into resonance. It is important to note that the
frequencies for transitions involving spin down depend on the
difference $\Delta_R-\Delta_L$ while the one involving spin up
does not. As we will show below, this is the main reason for
requiring an inhomogeneous magnetic field to obtain spin-up
polarized current. One should remark also that the resonance
achieved between
$|\uparrow\downarrow,\downarrow\rangle\Leftrightarrow|\uparrow,T_-\rangle$
occurs at a photon frequency:
$\omega_{T_-\downarrow}=\omega_{T_0\downarrow}$. Similarly, the
resonance
$|\uparrow\downarrow,\downarrow\rangle\Leftrightarrow|\downarrow,T_0\rangle$
occurs at $\omega_{T_0\uparrow}=\omega_{T_+\uparrow}$.

We have analyzed both homogeneous ($\Delta_R=\Delta_L$) and
inhomogeneous ($\Delta_R \ne \Delta_L$) magnetic field cases.\\
\\

\begin{figure}[htb]
\includegraphics[angle=270,width=3.5in,clip]{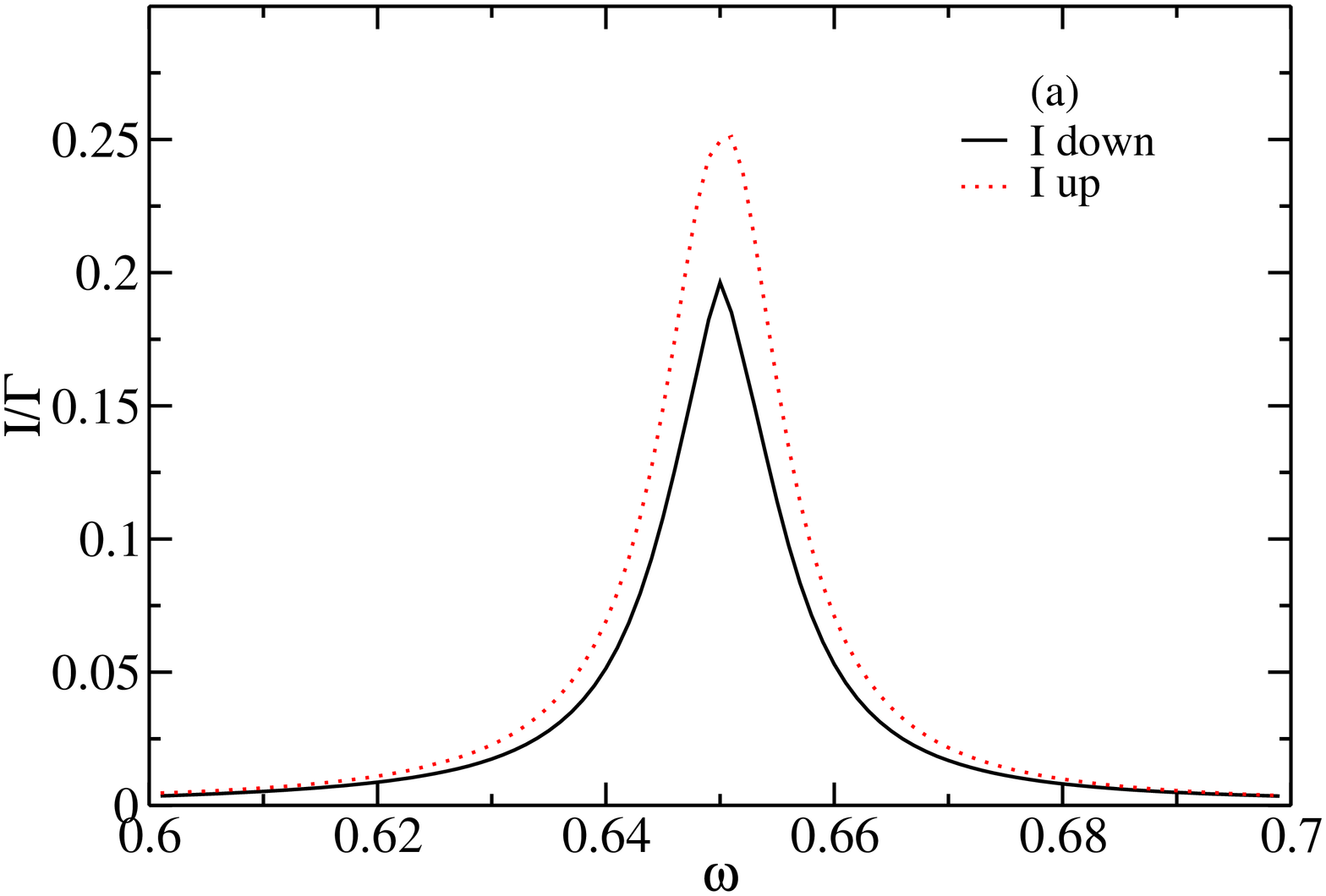}
\includegraphics[angle=270,width=3.5in,clip]{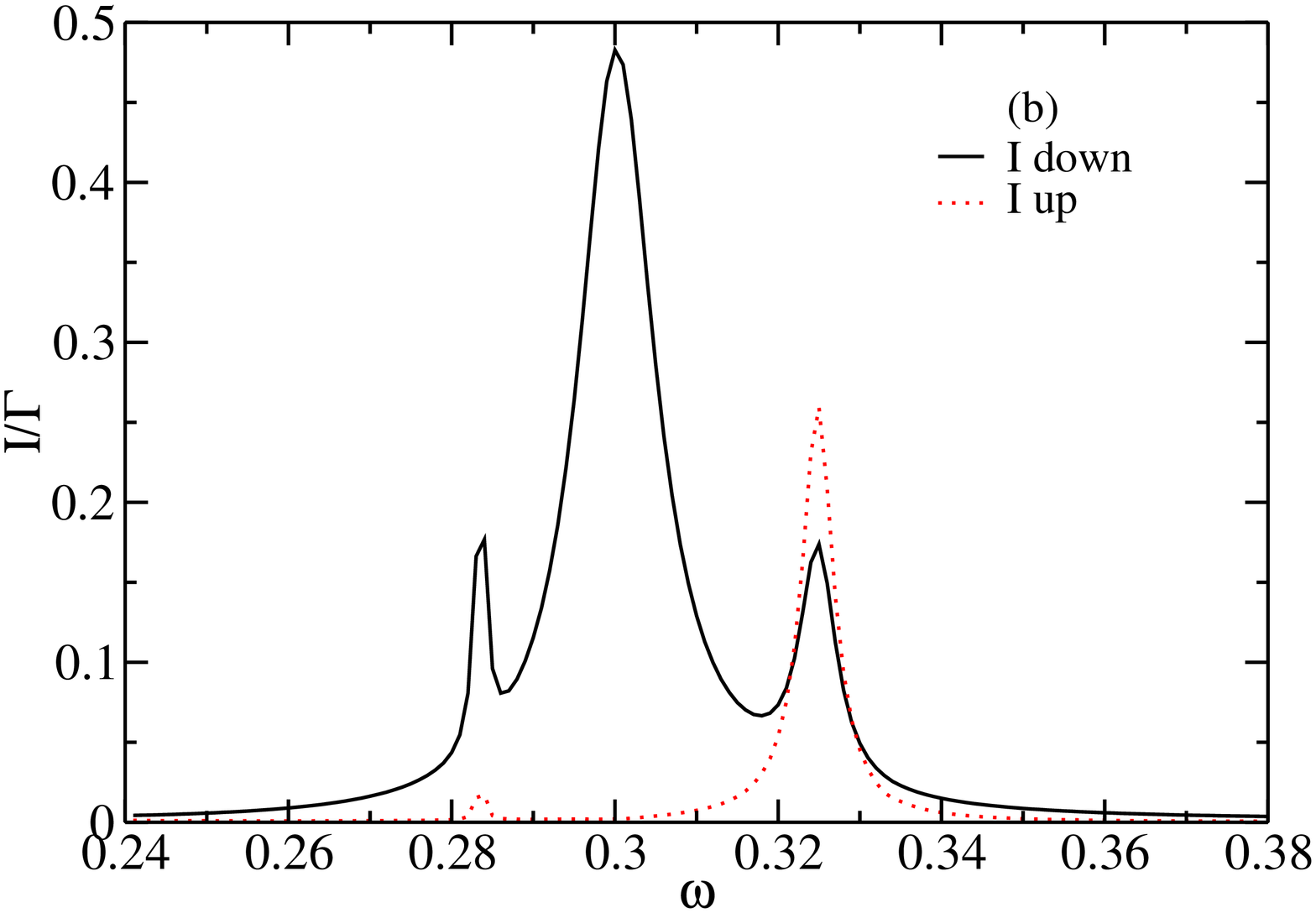}
\caption{\label{IvswDeltasignoTrnoCot}{\small Pumped current as a
function of the ac frequency ($\Delta_L=\Delta_R$), for the
resonances (a) $\omega_{T_0\downarrow}=\omega_{T_+\uparrow}\approx
0.65$ and (b) $\omega_{S_0\downarrow}\approx 0.3$. At this frequency, the
current is fully spin down polarized. The smaller
peaks in (b) are multi-photon satellites of other processes: at
$\omega \approx 0.283$, the three-photon process corresponding to
the resonance between the singlet $S_0$ in the left dot and the
singlet $S_1$ in the right dot occurs. The two overlapping peaks at $\omega
\approx 0.325$ correspond to two-photon satellites of the resonance in (a). Parameters:
$t_{LR}=0.005$, $\Gamma=0.001$, $U_L=1.0$, $U_R=1.3$,
$\Delta_L=\Delta_R=0.026$, $\Delta\varepsilon=0.4$, $J=-0.2$,
$\mu_L=\mu_R=1.31$, $V_{ac}=\omega_{T_+\uparrow}$.}}
\end{figure}

{\em i) $\Delta_R=\Delta_L$.}\\
\\

As has been discussed above, by tuning the ac frequency a series
of spin-polarized current peaks are obtained which can be
identified and related to different inter-dot resonant processes.
If the Zeeman splitting is the same in both dots, inter-dot
tunneling processes involving electrons with different spins, as
for instance:
$|\uparrow\downarrow,\uparrow\rangle\Leftrightarrow|\uparrow,T_0\rangle$
(at $\omega=\omega_{T_0\downarrow})$ and
$|\uparrow\downarrow,\uparrow\rangle\Leftrightarrow|\downarrow,T_+\rangle$
(at $\omega=\omega_{T_+\uparrow})$ occur at the same frequency,
i.e., $\omega_{T_0\downarrow}=\omega_{T_+\uparrow}\approx 0.65$
and we find two overlapping peaks
(Fig.\ref{IvswDeltasignoTrnoCot}(a)). The current is created
through several processes. The dominant ones are
$|\uparrow\downarrow,\uparrow\rangle\Leftrightarrow|\uparrow,T_0\rangle\rightarrow
\{|\uparrow,\uparrow\rangle
or|\uparrow\downarrow,T_0\rangle\}\rightarrow|\uparrow\downarrow,\uparrow\rangle$,
which contributes to spin-down current, and
$|\uparrow\downarrow,\uparrow\rangle\Leftrightarrow|\downarrow,T_+\rangle\rightarrow
\{|\downarrow,\uparrow\rangle
or|\uparrow\downarrow,T_+\rangle\}\rightarrow|\uparrow\downarrow,\uparrow\rangle$,
which contributes to spin-up current. So, in this case, the
current is partially spin-up polarized.

The frequency $\omega=\omega_{S_0\downarrow}\approx 0.3$ (Fig.
\ref{IvswDeltasignoTrnoCot}(b)), corresponds to the one-photon
resonance
$|\uparrow\downarrow,\uparrow\rangle\Leftrightarrow|\uparrow,\uparrow\downarrow\rangle$,
which is responsible of the large spin-down current peak through
the sequence:
$|\uparrow\downarrow,\uparrow\rangle\Leftrightarrow|\uparrow,\uparrow\downarrow\rangle\rightarrow
\left\{|\uparrow,\uparrow\rangle
or|\uparrow\downarrow,\uparrow\downarrow\rangle\right\}\rightarrow|\uparrow\downarrow,\uparrow\rangle$.
These are the only processes that take place at this frequency, so
no spin-up current is expected. Two peaks appear in the vicinity
of $\omega_{S_0\downarrow}$, one at
$\omega=\omega_{S_1\downarrow}/3\approx 0.283$ (which corresponds
to the three-photon satellite of resonance
$\omega_{S_1\downarrow}$ (not shown)) and two overlapping peaks at
$\omega \approx 0.325$ (corresponding to the two-photon satellites
at $\omega_{T_0\downarrow}/2=\omega_{T_+\uparrow}/2$, see
Fig.\ref{IvswDeltasignoTrnoCot}(a)). The positions of these
peaks are completely independent of each other and are determined
by the energetics of the system. Thus, we obtain fully spin-down
polarized current at $\omega=\omega_{S_0\downarrow}$, i.e., our
device acts as a filter for spin-down electrons.\\
\\

{\em ii) $\Delta_R \ne \Delta_L$.}\\
\\

\begin{figure}[htb]
\includegraphics[angle=270,width=3.5in,clip]{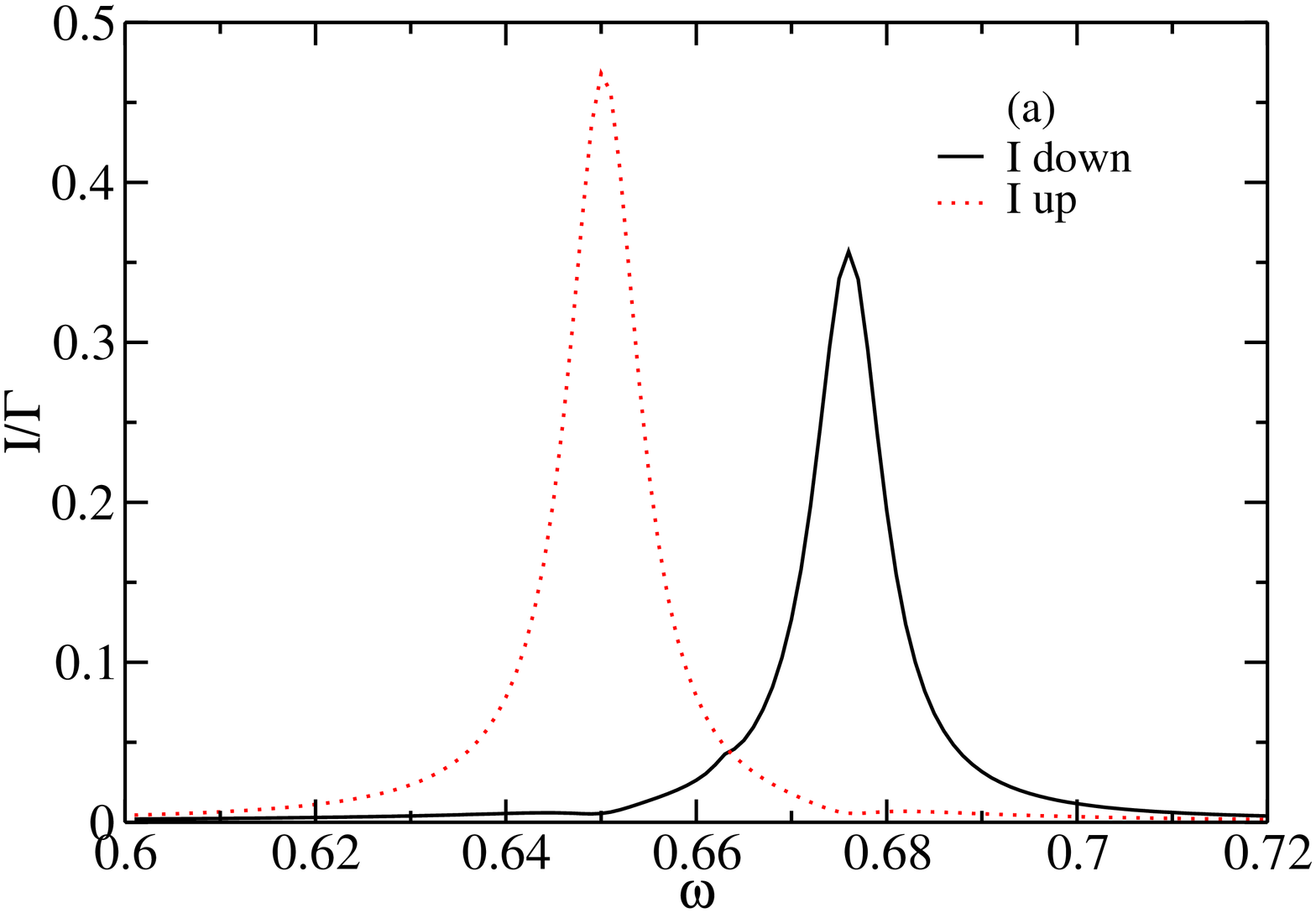}
\includegraphics[angle=270,width=3.5in,clip]{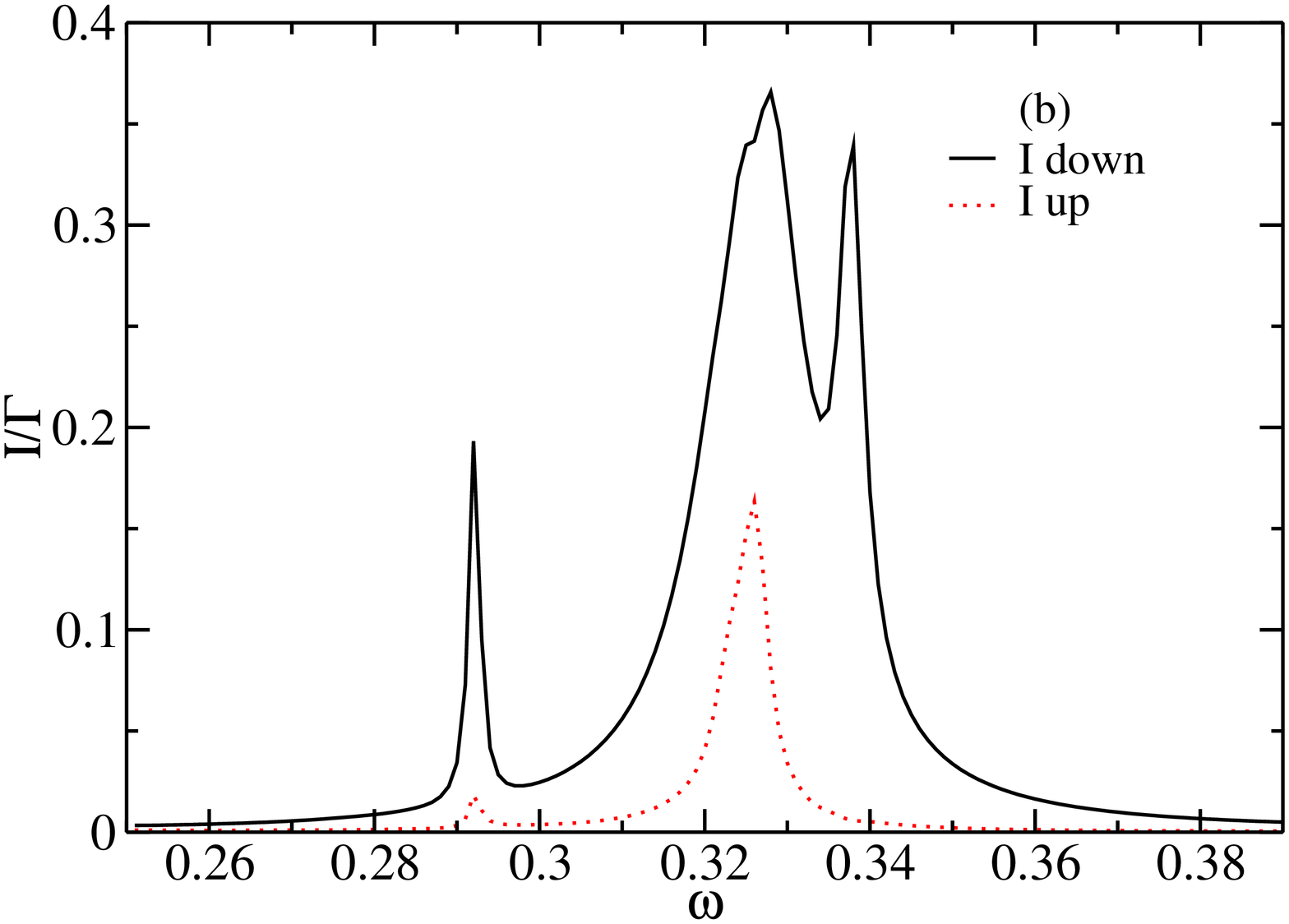}
\caption{\label{Ivsw0_5Delta2noTrnoCot} {\small Pumped current as
a function of the ac frequency ($\Delta_L\neq\Delta_R$), for the
resonances (a) $\omega_{T_0\uparrow}=\omega_{T_+\uparrow}\approx
0.65$ and
$\omega_{T_{0\downarrow}}=\omega_{T_{-\downarrow}}\approx 0.676$
and (b) $\omega_{S_{0\downarrow}}\approx 0.326$. The current at
the resonance $\omega=\omega_{S_0\downarrow}$ is partially spin
down polarized due to the overlapping satellite peaks (at
$\omega=\omega_{T_0\downarrow}/2\approx 0.338$,
$\omega=\omega_{T_+\uparrow}/2\approx 0.325$,
$\omega=\omega_{S_1\downarrow}/3\approx 0.292$). Parameters:
$t_{LR}=0.005$, $\Gamma=0.001$, $U_L=1.0$, $U_R=1.3$,
$\Delta_L=0.026$, $\Delta_R=2\Delta_L$, $\Delta\varepsilon=0.4$,
$J=-0.2$, $V_{ac}=\omega_{T_+\uparrow}$.}}
\end{figure}

In order to get a fully spin-up polarized current peak, we
consider different Zeeman splittings between both dots. This
introduces a separation
$\Delta\omega\approx\frac{\Delta_R-\Delta_L}{n}$ (where $n$ is the
number of photons involved in the resonant transition) between
peaks with different spin polarization, as can be seen for example
by comparing figures \ref{IvswDeltasignoTrnoCot}(a) and
\ref{Ivsw0_5Delta2noTrnoCot}(a).
Similarly, the resonances
$\omega_{T_0\downarrow}(=\omega_{T_{-}\downarrow}$)are shifted by
an amount $\Delta_R-\Delta_L$ with respect to the resonance at
$\omega_{T_+\uparrow}(=\omega_{T_0\uparrow}$) which is independent
of Zeeman splitting (see Fig.\ref{Ivsw0_5Delta2noTrnoCot}(a)). So
we obtain fully spin-up polarized current at
$\omega_{T_+\uparrow}\approx 0.65$ and fully spin-down polarized
current at $\omega_{T_0\downarrow}\approx 0.676$. In
Fig.\ref{Ivsw0_5Delta2noTrnoCot}(b), the resonance
$\omega=\omega_{S_0\downarrow}\approx 0.326$ is not well resolved
because there are overlapping satellite peaks (at
$\omega=\omega_{T_0\downarrow}/2\approx 0.338$,
$\omega=\omega_{T_+\uparrow}/2\approx 0.325$,
$\omega=\omega_{S_1\downarrow}/3\approx 0.292$). In concrete, we
find different processes that contribute to opposite spin
polarization currents and depend on the absorption of a different
number of photons (therefore, their Rabi frequencies are
renormalized with Bessel functions of different orders, Eq.
(\ref{Rabi})). It has been shown that for certain ac parameters
and sample configurations \cite{stafford}the height of the current
peaks depend on $\Omega_R$, which is a non linear function on the
ac intensity (see Eq. (\ref{Rabi})). Thus, we can manipulate the spin
polarization by tuning the intensity of the ac field.
\begin{figure}[htb]
\includegraphics[angle=270,width=3.5in,clip]{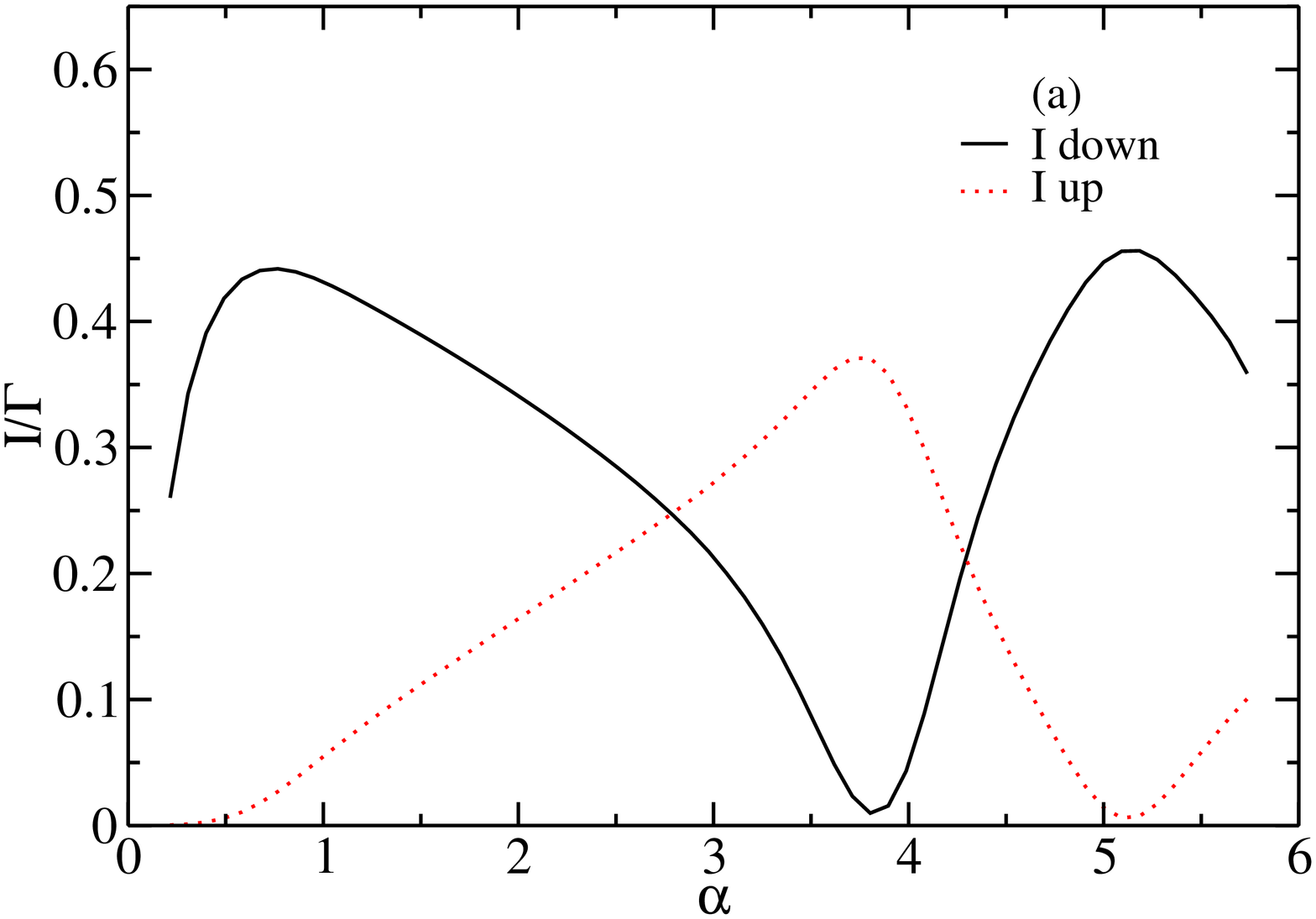}
\includegraphics[angle=270,width=3.5in,clip]{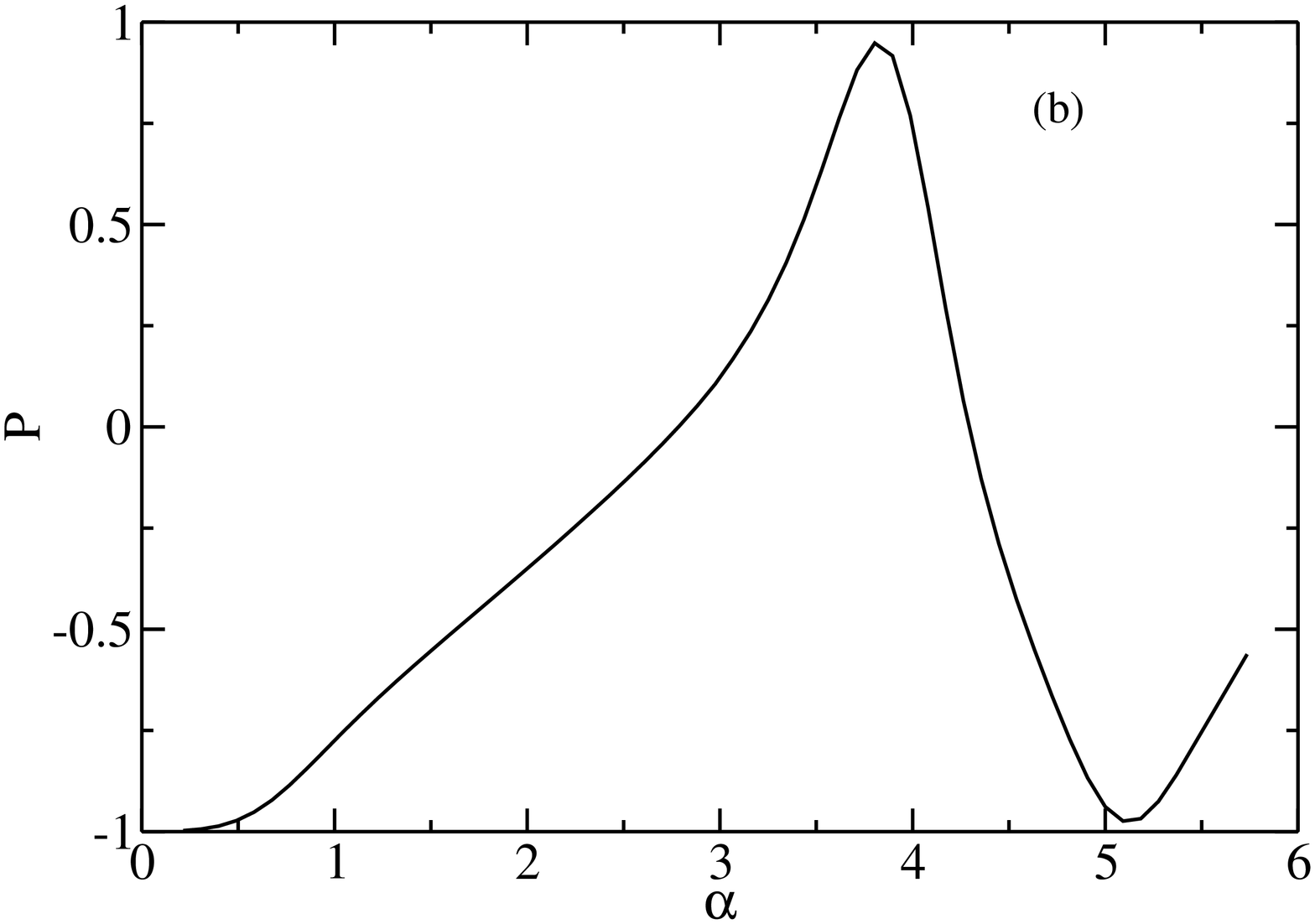}
\caption{\label{Ivsalfawdown05Delta1NotrNocot}{\small (a) Current
and (b)current polarization (defined as
$P=(I_{\uparrow}-I_{\downarrow})/(I_{\uparrow}+I_{\downarrow})$)
dependence on the ac field intensity, for
$\omega_{S_0\downarrow}=0.326$. There is fully spin-up
polarization for intensities such that $J_1(\alpha)=0$ and fully
spin-down-polarization when $J_2(\alpha)=0$. The sample parameters
are the same as in fig. 2.}}
\end{figure}
In Fig. \ref{Ivsalfawdown05Delta1NotrNocot} we show the current
and  current polarization as a function of ac-field intensity at
$\omega$=$\omega_{S_0\downarrow}$. Here, the spin-up contribution
comes from a two-photon resonance
($\omega=\omega_{T_+\uparrow}/2\approx 0.325$) and, thus, is
expected to vanish when $J_2(\alpha)=0$ due to the dynamical
localization phenomena\cite{report1}. Furthermore, for ac
intensities such that $J_1(\alpha)=0$, the one-photon resonance
$\omega_{S_0\downarrow}$ is quenched and we obtain at this frequency fully spin-up
polarized current. Thus, if we tune the ac intensity to the value
where the first(second)-order Bessel function vanishes, we obtain
fully spin-up(down) current.
Fig.\ref{Ivsalfawdown05Delta1NotrNocot}(a) shows that at low ac
intensities, the contribution of multi-photon processes is small
and the $\omega_{S_0\downarrow} \approx 0.326$ resonance
corresponding to practically fully spin-down current is clearly
resolved.

\section{Spin relaxation effects}

It is important to note also that, contrary to the case
for spin-down pumping, the pumping of spin up electrons leaves the
double dot in the excited state $|\downarrow,\uparrow\rangle$.
This makes the spin-up current sensitive to spin
relaxation processes.
If the spin $\downarrow$ decays before the
next electron enters into the left dot, a spin-down current
appears through the cycle
$(\downarrow\uparrow,\uparrow)\stackrel{\rm AC}\Leftrightarrow
(\downarrow,T_+)\stackrel{\Gamma_R}\Rightarrow
(\downarrow,\uparrow)\stackrel{W_{\uparrow\downarrow}}\Rightarrow
(\uparrow, \uparrow)\stackrel{\Gamma_L}\Rightarrow
(\uparrow\downarrow,\uparrow)$ and the pumping cycle is no longer
100\% spin-up polarized leading to a reduction of the spin-up current at this frequency.
\begin{figure}[htb]
\includegraphics[angle=270,width=4in,clip]{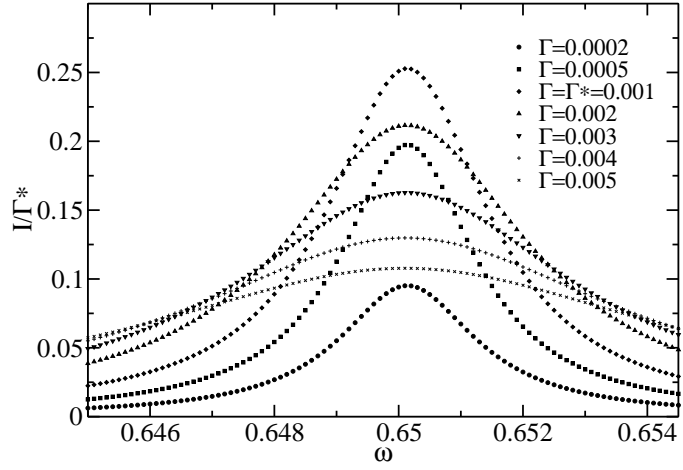}
\caption{\label{Ivsw0_1T1Vac014Gammas}{\small Pumped current near
resonance $\omega=0.65$ for different symmetric couplings to the
leads. $\Delta_z^R=\Delta_z^L/0.3$,
$W_{\uparrow\downarrow}=5\times 10^{-6}$, $V_{AC}=0.14$ and
${T_2}^*=0.1T_1$.}}
\end{figure}

\begin{figure}[htb]
\includegraphics[angle=270,width=3.5in,clip]{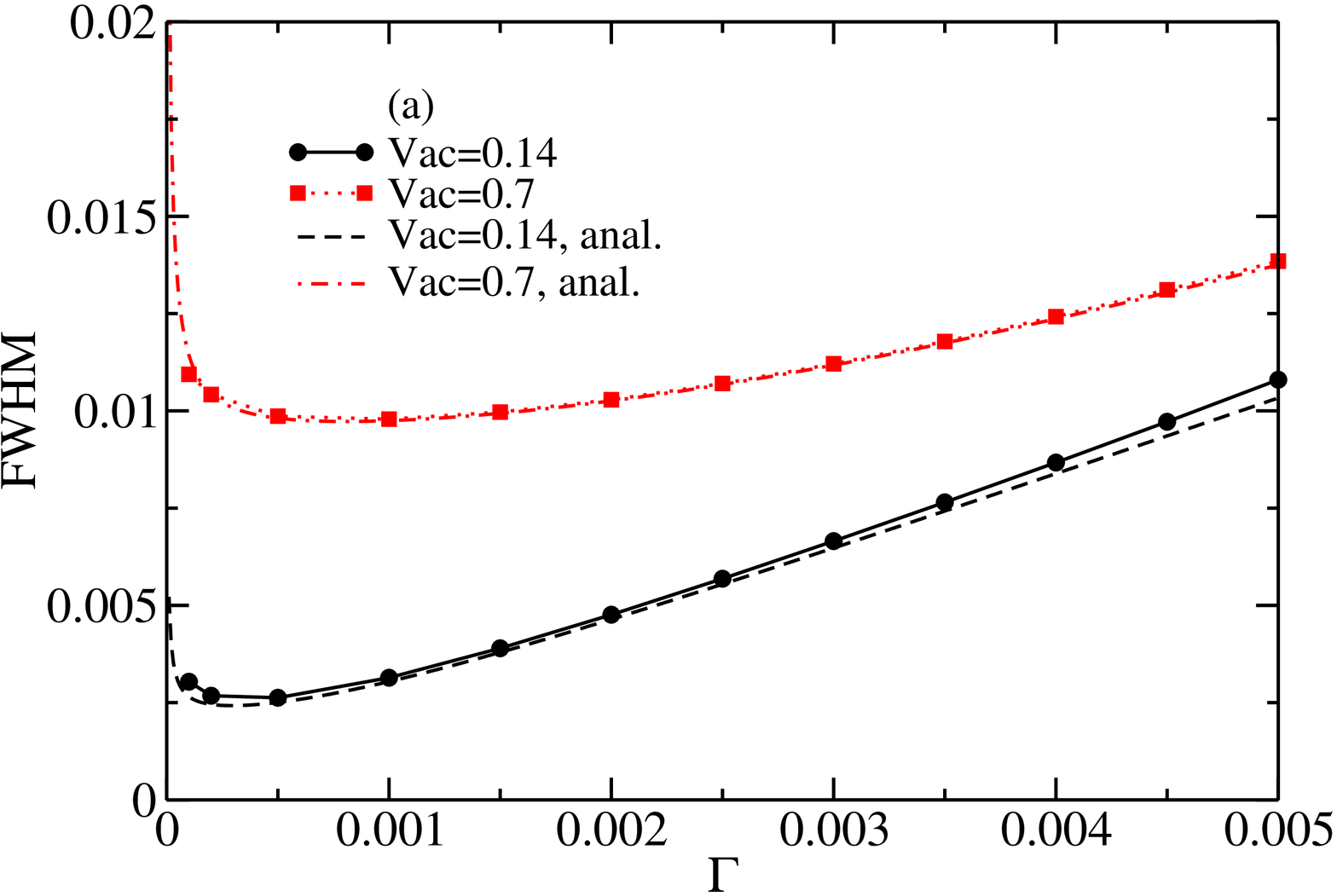}
\includegraphics[angle=270,width=3.5in,clip]{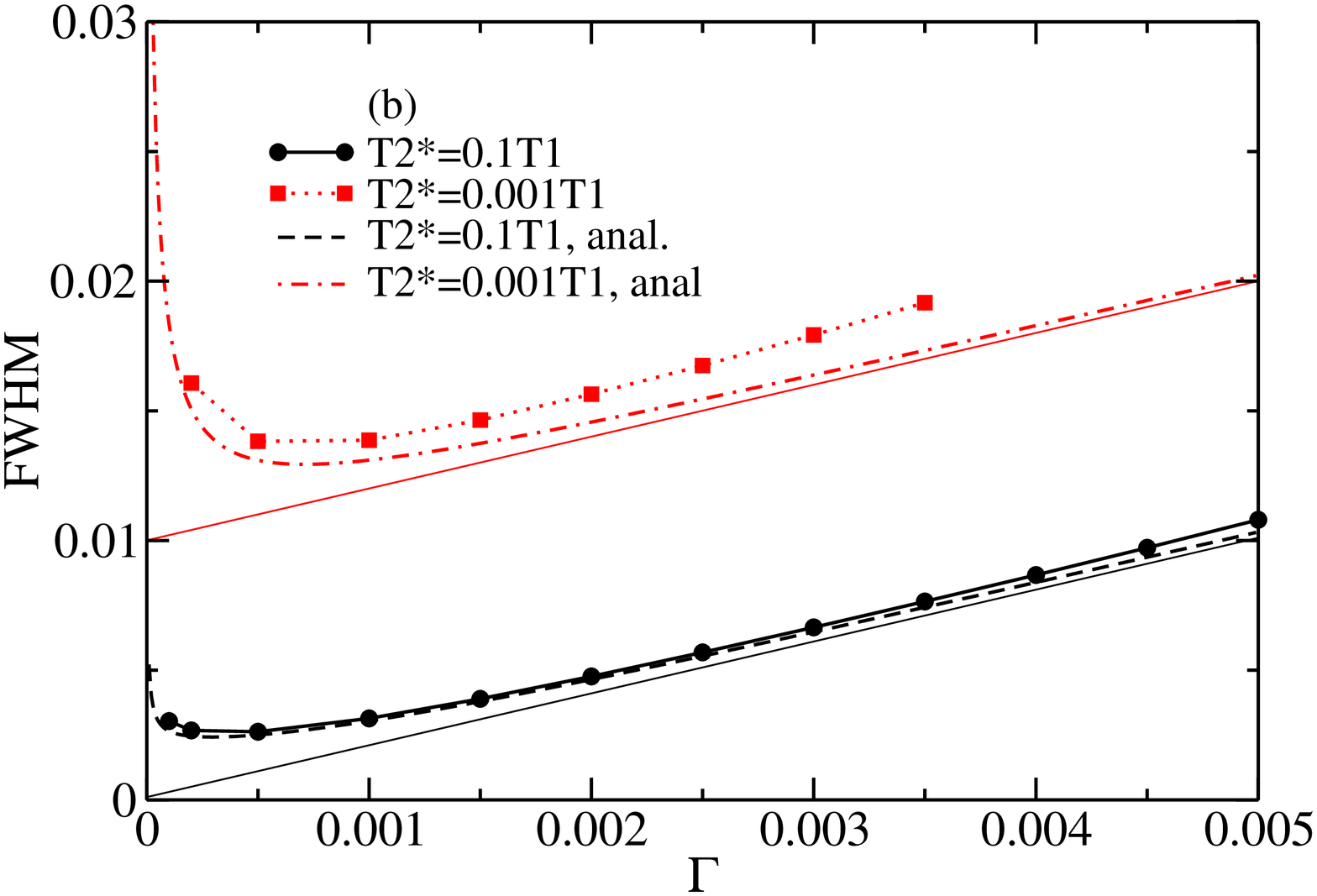}
\caption{\label{fwhmvsGammaWflip0_000005}{\small Full width at
half maximum of the total current in Fig.
\ref{Ivsw0_1T1Vac014Gammas} as a function of coupling to the
leads, $\Gamma$, at frequency $\omega=0.65$ (a) for strong
(squares) and weak (circles) field intensities and
${T_2}^*=0.1T_1$ and (b) for different spin dephasing times
${T_2}^*=.1T_1$ (circles) and ${T_2}^*=.001T_1$ (squares), for the
weak field intensity case. In both cases, the numerical results
are compared to the analytic prediction (dashed lines) given by
(\ref{width1}) . (a) In the weak field case, $V_{AC}=0.14$,
$\Omega_R \approx 0.001$ (circles), we see that for $\Gamma\gg
\Omega_R$, the curve follows a linear behavior: ${\rm FWHM}\approx
2\gamma=2(W_{\uparrow\downarrow}/2+1/{T_2}^*+\Gamma)$. For the
high field intensity case $V_{AC}$=0.7, $\Omega_R \approx 0.004$
and the same behavior is expected at larger $\Gamma$. In (b), for
${T_2}^*=.001T_1$ (squares), we see that the width of the peak is
larger than the expected asymptotic behavior for large $\Gamma$.
This is due to overlapping with the peak at
$\omega_{T_0\downarrow}$. As discussed in the text, the
extrapolation of the asymptotic curves at $\Gamma=0$ gives the
value of $2(1/2T_1+1/{T_2}^*)$. Thus, for the case where
${T_2}^*\ll T_1$, would allow to estimate the value of the spin
dephasing time. In both graphics, $\Delta_z^R=\Delta_z^L/0.3$,
$W_{\uparrow\downarrow}=1/T_1=5\times 10^{-6}$. }}
\end{figure}

We include a finite spin-flip relaxation probability,
$W_{\uparrow\downarrow}=1/T_1=5\times 10^{-6}{\rm meV}$ and
${T_2}^*$=0.1$T_1$ and calculate the current at
$\omega_{T_+\uparrow}=0.65$ for different values of the coupling with the
contacts, $\Gamma$, at $V_{AC}=0.14$ meV (Fig.
\ref{Ivsw0_1T1Vac014Gammas}).
The full widths (FWHM) of the current peaks
are plotted as a function of $\Gamma$ in Fig.
\ref{fwhmvsGammaWflip0_000005}(a) for weak (circles) and strong
(squares) AC-field intensity, $V_{AC}$.
 In order to minimize nonlinear effects, in Fig.
\ref{fwhmvsGammaWflip0_000005}(b) we investigate the low intensity
regime ($V_{AC}=0.14$) where we expect a FWHM dominated by
decoherence, for two different values of the spin dephasing time: 
${T_2}^*=0.1T_1$ and ${T_2}^*=.001T_1$. In every case, we find
that for large $\Gamma$, the behavior of the FWHM is linear with a
slope which approaches $2$, i.e., $\rm FWHM \sim 2 \Gamma$, which
can be directly related to the decoherence time $T_2$ as we show
below. Thus, experiments along these lines would complement the
information about decoherence extracted from other setups
\cite{Engel-Loss}. As an illustration
 we present below an analytical treatment which
allows us to relate the current peak widths with the spin
decoherence time.


\subsection{Analytical treatment}

In the following we present an analytical treatment in the stationary regime of the case discussed
above where the influence of spin-flip processes on the
spin up current peak coming from the ac-induced resonance between
$|1\rangle=|\uparrow\downarrow,\uparrow\rangle$ and
$|2\rangle=|\downarrow,\uparrow\uparrow^*\rangle$ was numerically obtained.

As discussed in the previous section, at $\omega=\omega_{T_+\uparrow}$, these states
are brought into resonance and the current, for $\Delta_R\approx \Delta_L/0.3$,
is fully spin-up polarized. Intermediate states
are $|3\rangle=|\downarrow,\uparrow\rangle$,
$|4\rangle=|\uparrow\downarrow,\uparrow\uparrow^*\rangle$,
$|5\rangle=|\uparrow,\uparrow\uparrow^*\rangle$ and $|6\rangle=|\uparrow,\uparrow\rangle$. From
\cite{stafford,hazelzet} it is known that the dynamics of the
system at large time scales is
  obtained by a time-dependent basis
transformation on the density matrix (rotating wave approximation,
RWA) such that
 $\epsilon(t)\rightarrow \epsilon_0 - n\omega$ and
$t_{LR}\rightarrow \tilde t_{LR}=(-1)^n J_n(V_{AC}/\omega)t_{LR}$. The
equations of motion for the corresponding reduced density matrix elements
in the RWA are:
\begin{eqnarray}
\dot\rho_{1}&=&-2\tilde t_{LR} \Im \rho_{2,1}+\Gamma_L\rho_3+\Gamma_R\rho_{4}+\Gamma_L\rho_6 \nonumber\\
\dot\rho_{2}&=&2\tilde t_{LR}\Im\rho_{2,1}-(\Gamma_R+\Gamma_L+W_{\uparrow\downarrow})\rho_{2}
\end{eqnarray}
while for the intermediate states,
\begin{eqnarray}
\dot\rho_3&=&\Gamma_R\rho_{2}-(\Gamma_L+W_{\uparrow\downarrow})\rho_3 \nonumber\\
\dot\rho_{4}&=&\Gamma_L\rho_{2}-\Gamma_R\rho_{4}+\Gamma_L\rho_{5} \nonumber\\
\dot\rho_{5}&=&W_{\uparrow\downarrow}\rho_{2}-(\Gamma_L+\Gamma_R)\rho_{5} \nonumber\\
\dot\rho_6&=&-\Gamma_L\rho_6+\Gamma_R\rho_{5}+W_{\uparrow\downarrow}\rho_3
\end{eqnarray}
The equation for the off-diagonal density matrix element is
\begin{equation}
\dot\rho_{2,1}=[i(\epsilon_0-n\omega)+\gamma]\rho_{2,1}+i\tilde t_{LR}(\rho_{2}-\rho_{1})
\end{equation}
where $\gamma=1/T_2$ is the decoherence rate:
\begin{equation}
\gamma=\frac{1}{2}(\Gamma_L+\Gamma_R+W_{\uparrow\downarrow})+\frac{1}{{T_2}^*}
\end{equation}
This, together with the condition of conservation of probability
$\Sigma_i \rho_i=1$ gives for the total current (at this
frequency), in the stationary regime, an expression which we can write\cite{hazelzet} as:
\begin{equation}
\label{lorentz}
 I=I_{0}\frac{W^2}{W^2+(\epsilon_0-n\omega)^2}
\end{equation}
where $I_{0}=2\gamma {\tilde t_{LR}}^2/W^2$ is the current maximum and W is the half width at half
maximum:
\begin{equation}
W^2=\frac{2\gamma {\tilde t_{LR}}^2}{\Gamma_L+\Gamma_R+W_{\uparrow\downarrow}}\left(\tilde\Gamma +
\frac{W_{\uparrow\downarrow}}{\Gamma_L+\Gamma_R}\left(\tilde\Gamma-1\right)\right) + \gamma^2
\end{equation}
Here, $\tilde
\Gamma=(\Gamma_L+\Gamma_R)^2/{\Gamma_L\Gamma_R}$.
For the symmetric case, $\Gamma_L=\Gamma_R=\Gamma$ ($\tilde\Gamma=4$) we can rewrite this in terms of the Rabi frequency
$\Omega_R=2{\tilde t_{LR}}$ as :
\begin{equation}
\label{width}
W^2=\frac{2\gamma\Omega_R^2}
{2\Gamma+W_{\uparrow\downarrow}}\left(1+\frac{3W_{\uparrow\downarrow}}{8\Gamma}\right)+\gamma^2
\end{equation}

In our calculations, we have taken $W_{\uparrow\downarrow}=5\times 10^{-6} \ll\Omega_R, \Gamma$ throughout the range of values considered.
Then, Eq. (\ref{width}) simplifies to
\begin{equation}
\label{width1}
W^2=\frac{\gamma\Omega_R^2}{\Gamma}+\gamma^2.
\end{equation}

 In the limit $W_{\uparrow\downarrow},1/T_2^* \ll \Gamma$, then
$\gamma \approx \Gamma$, and we get
$W^2=\Omega_R^2+\Gamma^2$ in agreement with previous analytical
results\cite{stafford,hazelzet}.

 From (\ref{width1}), we obtain the following asymptotic behaviors:
\begin{enumerate}
\item {$\Gamma\ll \Omega_R$ (strong inter-dot tunneling)$\Rightarrow
W^2\approx \Omega_R^2\gamma/\Gamma$
}
\item {$\Gamma\gg \Omega_R$ (weak inter-dot tunneling) $\Rightarrow
W\approx \gamma$}
\item {$\Gamma\approx \Omega_R \Rightarrow W^2\approx
\gamma_R(\Omega_R+\gamma_R)$,
where $\gamma_R=\Omega_R+1/{T_2}^*$.}
\end{enumerate}

In Fig. \ref{fwhmvsGammaWflip0_000005}(a) the full width at half
maximum (FWHM) of the spin-up current peak is represented as a
function of $\Gamma$ for the case ${T_2}^*=0.1T_1$ together with the analytical curve (\ref{width}).
For the weak field case $V_{AC}=0.14$ (full
circles) with $\Omega_R \approx 0.001$, we see that the
predictions of the theory are indeed fulfilled.
In particular, in
the range $\Gamma \gg \Omega_R$, the
 FWHM as a function of $\Gamma$ is a straight line with slope
2 as expected (${\rm FWHM}\approx
2\gamma=2/{T_{2}}^*+1/T_1+2\Gamma$). This means that a direct
measure of the decoherence time for the isolated system
$T_2^{iso}$:  $1/T_2^{iso}$=$1/{T_{2}}^*+1/2T_1$ can be obtained
from this linear behavior.
 From Fig. \ref{fwhmvsGammaWflip0_000005}(a) it can be verified
that the cases (i) ($W^2\approx a+b/(\Gamma T_2^{iso})$) and (iii) are
also reproduced. Besides, it is interesting to mention that from
these cases one can get information on the Rabi frequency. The
same analysis also holds for the strong field case $V_{AC}=0.7$
(full squares), where $\Omega_R \approx 0.004$.

Recently, it has been measured that the spin dephasing time
${T_2}^*$ induced by hyperfine interaction is tens of nanoseconds\cite{koppens,pettaScience}. 
This, together with experimental values
for $T_1$ as long as miliseconds\cite{hanson3} in GaAs quantum dots, such
that $1/T_1 \ll 1/{T_2}^*$,
would allow to estimate the spin dephasing time ${T_2}^*$ directly
from the intersection of the large $\Gamma$ asymptote with the
vertical axis (Fig. \ref{fwhmvsGammaWflip0_000005}(b), dotted
line).

We note, from  Fig. \ref{fwhmvsGammaWflip0_000005}(b), that the
numerical results differ from the expected analytical curves for
the case ${T_2}^*=10^{-3}T_1$ for large $\Gamma$. This is because
the parameters that contribute to FWHM ($\Gamma$ or ${1/T_2}^*$)
are large enough to make the current peak overlap with its
neighboring spin-down current peak at $\omega_{T_0\downarrow}$.
Then, it loses its lorenztian shape, mixes its spin polarization
and gets wider than what it is analytically expected. This problem
should hold in the experimental measurements unless the energy
difference between both peaks were large enough so they do not
overlap and they can be fitted to a lorenztian curve.
This is the case when $\Delta_R-\Delta_L\gg 2\gamma$.


\section {Summary}

We have proposed and analyzed a new scheme of realizing {\it both}
spin filtering and spin pumping by using ac-driven double quantum
dots coupled to unpolarized leads. Our results demonstrate that
the spin polarization of the current can be manipulated (including
fully reversing) by tuning the parameters of the ac field. For
homogeneous magnetic field, $\Delta_L=\Delta_R$, we obtain spin
down polarized current involving singlet states in both dots. In
order to obtain spin up polarized current, an inhomogeneous
magnetic field is required to break the degeneracy in transitions
involving triplet states in the right dot. Our results also show,
both analytically and numerically, that the width in frequency of
the spin-up pumped current gives information about the {\it spin
decoherence time} $T_2$ and also about the {\it spin dephasing
time} ${T_2}^*$ of  the isolated double quantum dot system.
Experiments along these lines
would allow to get
information, from transport measurements, on the different
mechanisms producing spin decoherence in quantum dots.
\\
\\
We thank J.A. Maytorena for useful discussions. Work supported
by Programa de Cooperaci\'on Bilateral CSIC-CONACYT, by the EU
grant HPRN-CT-2000-00144 and by the Ministerio de Ciencia y
Tecnolog\'{\i}a of Spain through grant MAT2002-02465.\\

\end{document}